\documentclass[12pt,a4paper]{article}
\usepackage[dvipdfm]{graphicx}
\usepackage{amsmath,amssymb}
\usepackage{amsfonts}
\usepackage{cite}
\def\be{\begin{equation}}
\def\ee{\end{equation}}
\def\bea{\begin{eqnarray}}
\def\eea{\end{eqnarray}}

\begin{document}
\begin{center}

 {\bf Effect of Tensor Correlations on Single-particle and Collective States}

 \medskip
 H. Sagawa$^{1}$

 \medskip
{\it
 $^{1}$Center for Mathematics and Physics, University of Aizu\\
Aizu-Wakamatsu, Fukushima 965-8580 , Japan

%% $^{2}$Affiliation-2
}
\end{center}

\vspace*{-1cm}
\begin{abstract}
We study the effect of tensor correlations on single-particle and 
collective states within Skyrme Hartree-Fock and RPA model.
Firstly, We study the role of tensor interactions in Skyrme effective 
interaction on the spin-orbit splittings of N=82 isotones and Z=50 isotope.
%%We analyze also the spin-orbit splittings 
%%in the Ca isotopes and in the N=28 isotones.
%%We analyze the evolution of the spin-orbit splittings 
%%in the Ca isotopes and in the N=28 isotones. 
%%We also focus on the 
%%reduction of the spin-orbit splittings associated with f and p
%%orbits from $^{48}$Ca to $^{46}$Ar. 
The isospin dependence 
of the shell structure is well described as the results of
the tensor interactions without destroying good properties of
the binding energy and the rms charge radii of the heavy nuclei. 
%%We conclude that adding the tensor 
%%contribution can  explain in most cases the empirical trends,
%% whereas  this is not the case if one simply employs existing Skyrme
%% parameterizations without the tensor forces~\cite{Colo07}. 
 Secondly, We performed self-consistent 
HF+RPA calculations for charge exchange
1$^+$ states in $^{90}$Zr  and $^{208}$Pb to elucidate the role of 
tensor interactions on spin dependent excitations.  
%% We have employed a parameter set in
%% which the tensor terms are added to the SGII interaction~\cite{Bai09}.
 It is pointed
out that Gamow-Teller(GT) states can couple  strongly with the
spin-quadrupole (SQ) 1$^+$ states in the high energy region above
E$_x$=30 MeV due to the tensor interactions. As the result of this
coupling, more than $10\%$ of the GT strength is shifted to the
energy region above 30 MeV, and the main GT peak is moved
2 MeV downward.
%%At the same time, the main SQ 1$^+$ peak is
%%moved upward by more than 10 MeV due to the tensor correlations.
%% The effect of tensor interactions on nuclear structure problems 
 %%will be  further discussed.  
  \end{abstract}

\section{Introduction}

\begin{figure}[htp*]
\centering
\includegraphics[width=12cm,clip]{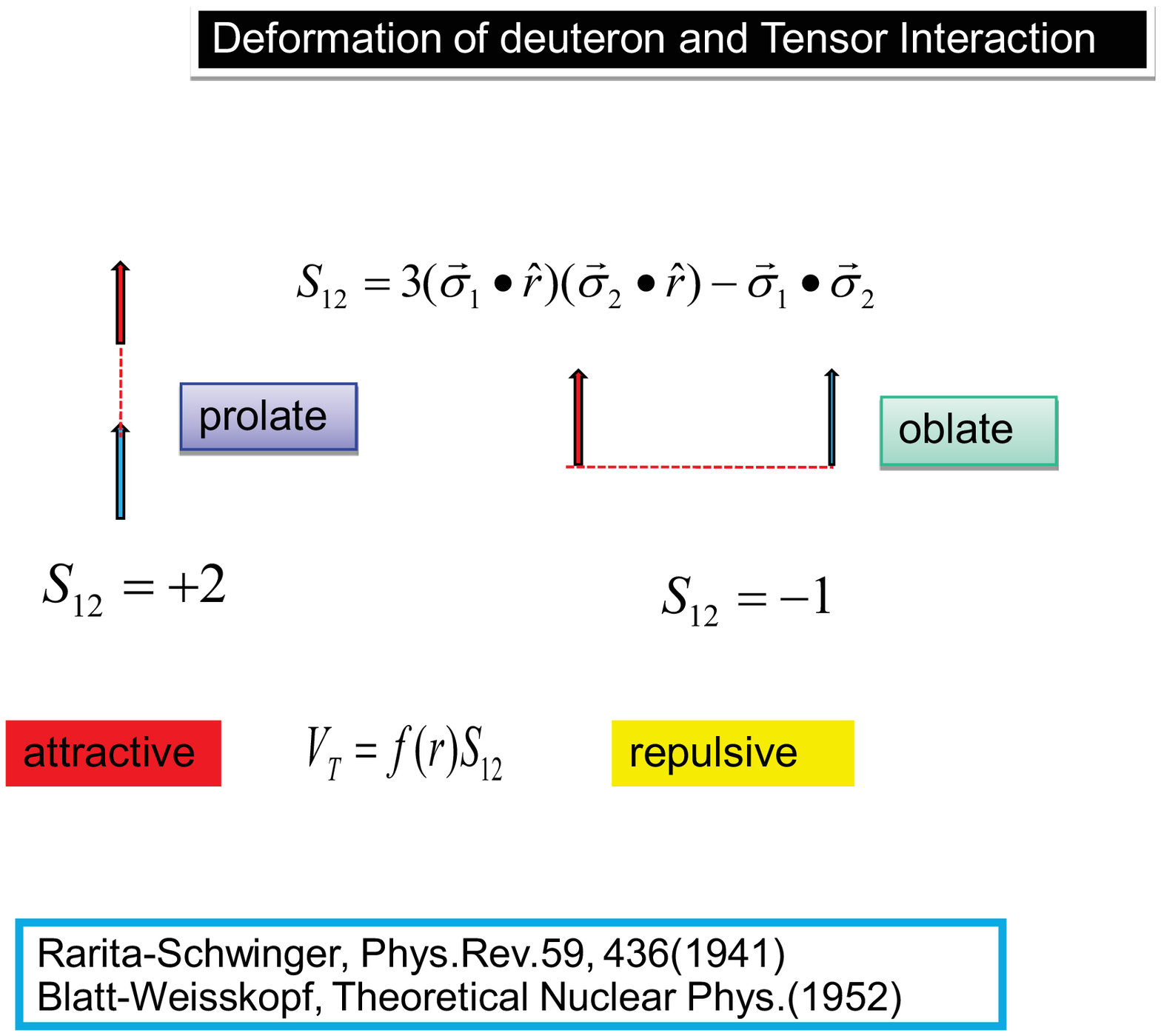}
\vspace*{-1cm}
\caption{\label{fig:a1}
%\begin{description}
%\item[{\rm Figure 1:}]
Tensor interaction.
See the text for details.
}
\end{figure}

%%% Nuclei far from the stability lines open  a new test ground for nuclear 
%%% models.  
%%% Recently, many experimental and theoretical efforts have been
%%% paid to study structure and reaction mechanism in nuclei near drip
%%% lines.   Modern  radioactive nuclear beams  and experimental detectors 
%%%  reveal several
%%% unexpected
%%%   structure of  nuclei such as
%%%  existence of halo and skins~\cite{Tani},
%%%    modifications of  shell closures~\cite{Ozawa}
%%% and Pigmy resonances in electric dipole transitions ~\cite{Pigmy}.
One of the current topics in nuclear physics 
 is  the role of the tensor interactions on the shell evolution 
 of single-particle states and  on spin-isospin excitations.
%%%  of nuclei far from the stability line.
 The importance of tensor interactions on nuclear many-body problems 
was recognized more that 50 years ago. Especially, it plays the essential 
role to make the binding systems such as the deuteron.
 As is shown Fig. \ref{fig:a1}, the tensor
 interaction acts on the spin triplet state (S=1) of two nucleon system.
%%deuteron can make the binding system 
%% only when the tensor interactions are introduced and it becomes 
%% strongly 
%% deformed   by the tensor interaction.  
When a proton  and a neutron are 
aligned in the direction of spins, the deuteron gets an extra binding energy 
by the tensor interaction, 
\be 
 V_T=f(r)S_{12}
\ee
since  $f(r)$ is negative and $S_{12}=3(\vec{\sigma}_1\cdot\hat{r})
(\vec{\sigma}_2\cdot\hat{r})-\vec{\sigma}_1\cdot\vec{\sigma}_2$=2.
On the other hand, if a proton and a neutron are  perpendicular to 
the spin direction, the deuteron will loose the binding energy since 
 $S_{12}=-1$  and
 cannot make a bound state.   
Thus, the deuteron becomes a strongly deformed 
  prolate shape  and makes a binding
 system only when the tensor correlations are taken into account.  
 
 After these findings, the importance of tensor interactions has been 
widely recognized in nuclear many-body systems, especially in light 
nuclei.  
The role of the tensor interactions  in the Hartree
-Fock calculations was firstly discussed by Stancu et al., thirty years ago
 \cite{Stancu}. 
However serious attempts have never been performed until very recently
 \cite{Colo07,Brown,Brink,Lesinski,Wei}.
The importance of the tensor correlations on the mean field revived 
by the study of the shell evolution of heavy exotic nuclei \cite{Otsuka}.
In this paper, I will summarize formulas of Skyrme tensor interactions
in Section 2.  The isotope dependence of the shell structure
of Z=50 isotopes and N=82 isotones are studied 
by using the Hartree-Fock (HF)+BCS 
calculations  in Section 3.  The role of tensor 
interactions on spin-isospin excitations 
 will be examined in Section 4. Summary is given in Section 5.
%%%%%% 12pm  Oct 23  %%%%%%

\section{Skyrme tensor interaction}
The tensor force was considered in the
Skyrme-Landau parameterization and the sum rules of electromagnetic 
transitions in Ref.~\cite{Liu}. However, the tensor force was
essentially dropped in most Skyrme parameter sets which have been
 used  widely in nuclear structure
calculations. Recently, in Ref.~\cite{Brown}, a Skyrme interaction
 was fitted including  the tensor contribution. Then, tensor
terms were added perturbatively in Refs.~\cite{Colo07} and~\cite{Brink}
  to the existing standard parameterizations
SIII~\cite{Beiner} and SLy5~\cite{Chabanat}, respectively.
Eventually, several new parameter sets have been fitted in
Ref.~\cite{Lesinski} and used for systematic investigations within
the Hartree-Fock-Bogoliubov (HFB) framework. The inclusion of tensor
terms in the Skyrme HF calculations achieved considerable success in
explaining some features of the evolution of single-particle
states~\cite{Colo07,Wei}. The Skyrme tensor interaction is given by 
   the triplet-even and triplet-odd tensor zero-range tensor parts,
\begin{eqnarray}
v_T &=& {T\over 2} \{ [ (\sigma_1\cdot k^\prime)
(\sigma_2\cdot k^\prime) - {1\over 3}(\sigma_1\cdot\sigma_2)
k^{\prime 2} ] \delta(r_1-r_2) 
\nonumber \\
&+& \delta(r_1-r_2)
[ (\sigma_1\cdot k)(\sigma_2\cdot k) - {1\over 3}
(\sigma_1\cdot\sigma_2) k^2 ] \} 
\nonumber\\
&+& U \{ (\sigma_1\cdot k^\prime) \delta(r_1-r_2) 
(\sigma_1\cdot k) - {1\over 3} (\sigma_1\cdot\sigma_2) 
[k^\prime\cdot \delta(r_2-r_2) k] \}  
\label{eq:tensor}
\end{eqnarray}
where the operator ${\bf k}=(\nabla_1-\nabla_2)/2i$ acts on the right and
${\bf k}'=-(\nabla_1-\nabla_2)/2i$ on the left. 
The coupling constants $T$ and $U$ denote the strength of the 
 triplet-even and triplet-odd tensor interactions, respectively. 
We treat these coupling constants as free parameters in the following study. 
The tensor interactions (\ref{eq:tensor}) give the contributions to the 
binding energy and the spin-orbit splitting proportional to the spin density
\begin{equation}
 J_q(r)=\frac{1}{4\pi r^3}\sum_{i}v_{i}^2(2j_{i}+1)\left[j_i(j_i+1)
           -l_i(l_i+1) -\frac{3}{4}\right]R_i^2(r)
\label{eq:sd}
\end{equation}
where $i=n,l,j$ runs over all  states and $q=0 (1)$ is the 
isospin quantum number for neutrons (protons).  
The $v_{i}^2$ is the occupation probability 
of each orbit determined by the BCS approximation and $R_i(r)$ is 
the HF single-particle wave function.  It should be noticed that 
the exchange part of the central Skyrme interaction gives  
the same kind of contributions to the total energy density.
The central exchange and tensor
 contributions give the extra terms to the energy density as 
\begin{equation}
\delta E= {1\over 2}\alpha(J_n^2+J_p^2) + \beta J_n J_p.
\label{eq:dE}
\end{equation}
The spin-orbit potential is then expressed to be
%\begin{eqn}
\bea
U_{s.o.}^{(q)} = {W_0\over 2r} \left( 2{d\rho_q\over dr} + 
{d\rho_{1-q}\over dr} \right) + \left( \alpha 
{J_q\over r} + \beta {J_{1-q}\over r} \right). 
\label{eq:dW}
\eea
%\end{eqn}
where the first term on the r.h.s comes from the  Skyrme 
spin-orbit interaction and the second term include both the central
 exchange the tensor contributions $\alpha= \alpha_c +\alpha_T$ 
and $\beta=\beta_c +\beta_T$.  In Eq. \eqref{eq:dW}, 
$q=0 (1)$ is assigned for neutrons (protons).
 The central exchange contributions  are
given by 
\begin{eqnarray}
\alpha_C & = & {1\over 8}(t_1-t_2) - {1\over 8}
(t_1x_1+t_2x_2) \nonumber\\
\beta_C & = & -{1\over 8}(t_1x_1+t_2x_2).  
\label{eq:dWc}
\end{eqnarray}
where the parameters are defined in ref. \cite{Skyrme}.
The tensor contribution are expressed as 
\begin{eqnarray}
\alpha_T & = & {5\over 12}U \hfill\nonumber\\
\beta_T & = & {5\over 24}(T+U).
\label{eq:dWT}
\end{eqnarray}
%%%The central exchange contributions (\ref{eq:dWc})
%%%  have been neglected in most of 
%%%previous HF calculations of Skyrme interactions.  
%%% In order to have a general idea 
%of the 
%contributions
%%% of the central and tensor interactions to the 
%%%spin-orbit splitting, we include the central term  (\ref{eq:dWc}) 
%%%in the following calculations.  For another  positive reason,
%%% we should keep the two
%%% terms (\ref{eq:dWc}) and (\ref{eq:dWT}) in the HF calculations
%%%for  future fully self-consistent 
%%%HF+ random phase approximation (RPA) calculations including the tensor
 %%%interactions.

\begin{figure}[ht]
\centering
\includegraphics[width=12cm,clip]{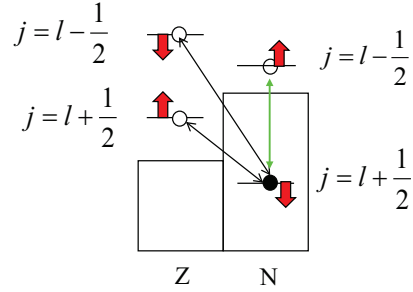}
\vspace*{-2.5cm}
\caption{\label{fig:a2}
%\begin{description}
%\item[{\rm Figure 2:}]
Tensor and Spin-orbit interaction.
See the text for details.
}
\end{figure}

Before going to detailed study, let us 
mention two important features of the tensor and the central exchange 
contributions in Eq. (\ref{eq:dW}) to the 
spin-orbit splitting.  
First point is that the mass number dependence of the  
the first and second terms in Eq.(\ref{eq:dW}). 
Since the Skyrme spin-orbit force $W_0$ gives the spin-orbit splitting
proportional to the derivatives of the densities, the mass number dependence
is very modulate in heavy nuclei.  On the other hand, the second term 
in Eq. (\ref{eq:dW}) depends on the spin density $J_q$ which has 
essentially no contribution
 for the $l\cdot$s closed shell.  The spin density will increase exactly 
 proportional to the number of particles in the open shells if one of the
spin-orbit partner is only active.
 Moreover, the sign of the $J_q$ will change depending upon 
 which orbits are  involved 
 in the active shell orbits, i.e., the orbit $j_{>}=l+1/2$ gives  a positive
  $J_q$ value while the orbit $j_{<}=l-1/2$ gives a negative  $J_q$.
  This means that the spin-orbit energy will change in the opposite 
  direction according to which orbit is occupied in  the open shell 
  nuclei. 

%% According to ref. \cite{Stancu}, the optimal parameters 
%% $\alpha_T$ and $\beta_T$ are shown in a triangle in the two dimensional 
%% ($\alpha_T$,  $\beta_T$) plane. 
  We fit the two parameters $T$ and $U$ (equivalently  $\alpha_T$ and $\beta_T$)
  using the recent 
 experimental data of $N=$82 isotones and $Z=50$ isotopes. We keep
the central part of Skyrme interaction as that of SLy5.
The central exchange interactions are 
 $\alpha_c$=80.2MeV$\cdot$fm$^5$ and $\beta_c=-$48.9
 MeV$\cdot$fm$^5$ for SLy5. 
 The optimal parameters  $\alpha_T$ and $\beta_T$ are determined to be
 ($\alpha_T$, $\beta_T)=(-$170,100)MeV$\cdot$fm$^5$.  
We examine detailed  properties of the tensor interactions
by using our parameter sets.  In Fig. \ref{fig:a2},
we consider a nucleus where the last occupied orbit is the 
neutron $j_>=l+1/2$  
(for example, imagine $^{90}$Zr or $^{48}$Ca).  
 In. Eq. \eqref{eq:dW}, the  $\alpha_T$ increase the spin-orbit splitting 
because of positive $J_{q=0}$ of $j_>=l+1/2$ orbit, 
while  $\beta_T$ has the  opposite sigh and decrease the splitting for 
positive  $J_{q=0}$. 
These effects are
  illustrated in  Fig. \ref{fig:a2} where the neutron spin-orbit 
 splitting is increased, while the proton one is decreased.

%%  We leave the optimum  refit of  all the 
%%Skyrme parameters including the tensor terms for 
%% future works.
%% Instead we show how much the binding energies  and the rms radii of 
%% $^{132}$Sn and $^{208}$Pb will be affected by the inclusion of the
%% tensor and the central  terms $\alpha$ and $\beta$ in the energy 
%% density.

\begin{figure}[htb]
\centering
\includegraphics[width=12cm,clip]{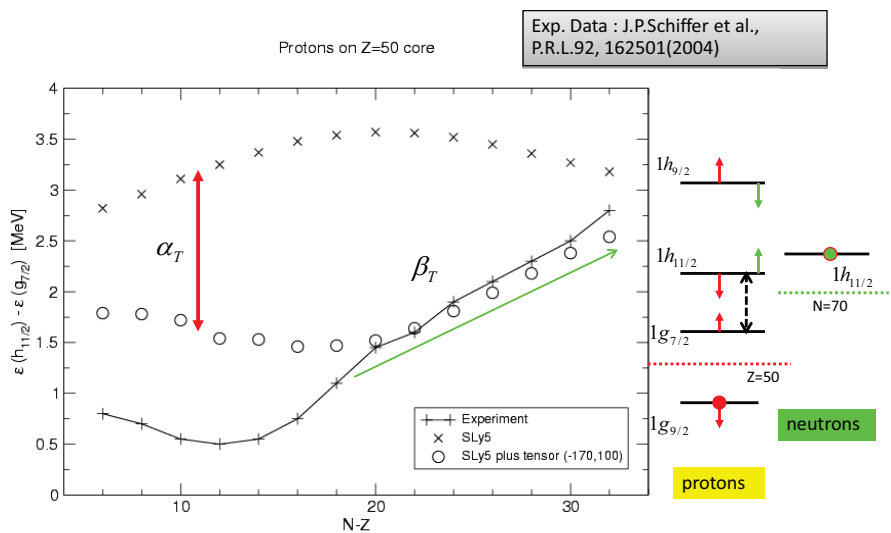}
\caption{\label{fig:Z50}
%\begin{description}
%\item[{\rm Figure 4 :}]
Comparison of energy difference between pairs of
 single-particle 1g$_{7/2}$ and 1h$_{11/2}$ proton states of Z=50 isotones.
The calculations are performed with and without tensor and central exchange
terms in the spin-orbit potential (\ref{eq:dW}) top of SLy5 parameter set.
The experimental data are taken from ref. \cite{Schiffer}.
See the text for details.
}
\end{figure}

In Fig. \ref{fig:Z50},  the energy differences for 
proton single-particle states $\Delta e(h_{11/2}-g_{7/2})$ of Z=50 isotones
 are shown 
as a function of the neutron excess $(N-Z)$.  The original SLy5 
interaction fails to reproduce the experimental trend qualitatively and
quantitatively.  Firstly, the energy differences of the HF results are 
much larger than the empirical data.  Secondly, 
the experimental data decrease, the neutron excess decreases and reach
about 0.5MeV at the minimum value. On the other hand,  the energy 
differences of the original 
SLy5 increase as the neutron excess decrease and has 
 the maximum at around $(N-Z)$=20.  We studied also 
several other Skyrme parameter sets and found almost the same 
trends as those of SLy5.  

The tensor  
 central exchange interactions are included in the results marked 
  by open circles in Fig. \ref{fig:Z50}.  
 We can see a substantial improvement by introducing the 
 tensor interactions.  The 
 set ($\alpha_T$, $\beta_T)=(-$170,100)MeV$\cdot$fm$^5$
   gives a fine agreement with 
 the empirical data from $(N-Z)=20\sim32$ quantitatively and qualitatively.
%% Namely, the results with the contributions $\alpha$ and $\beta$ 
%% decrease when the neutron excess decreases   to  $(N-Z)=16$

The HF+tensor results can be  qualitatively understood by
  general argument as follows.
  Firstly, the strength $\alpha$ change the depth of
  the proton spin orbit potential. For Z=50 core, only the  proton
  g$_{9/2}$  proton orbit dominates  the  spin density $J_p$ in Eq.
  (\ref{eq:sd})   so that with the negative $\alpha_T$= value
   the spin-orbit splittings  are  increased. 
Thus, the $g_{9/2}$ protons increase the spin-orbit splitting between 
 proton $(g_{9/2}-g_{7/2})$ orbits and that of proton $h_{13/2}-h_{11/2}$ 
 orbits by $\alpha_T$ effect.  As a net effect, 
 the $\Delta e(h_{11/2}-g_{7/2})$ protons decreases substantially. 
%%  Then energy difference 
%%  $\Delta$ e$(h_{11/2}-g_{7/2})$ is smaller for larger  negative $\alpha_T$. 
%   Notice that the $\alpha$ gives no isospin dependence on the
%  spin-orbit potential in general. 

  Next  let  us study the (N-Z) dependence where  $\beta_T=$ plays the 
 essential role.
 In the case of $\Delta e(h_{11/2}-g_{7/2})$ protons on Z=50 core
  from N-Z=(6-14), the $g_{7/2}$ neutron orbit is gradually filled.
  Then the $\beta_T=$100MeV$\cdot$fm$^5$ gives a negative contribution
  to the spin-orbit potential (\ref{eq:dW})
  and  increases the spin-orbit  splitting.  Therefore the
   energy difference $\Delta e(h_{11/2}-g_{7/2})$ is further decreasing.
  From  N$-$Z=(14 to 20), the 
  $s_{1/2}$ and $d_{3/2}$ neutron orbits are occupied.
  In this region the spin density is not so much changed since 
  $s_{1/2}$ has zero contribution.
  For N-Z=(20-32), the h11/2 orbit is  gradually filled. Then 
  this orbit gives a positive contribution to 
  the spin-orbit potential  (\ref{eq:dW}) 
 ,i.e., the the spin-orbit  splitting is decreasing.
  Then the $\Delta e$ turns out to be  increasing.  
  The magnitude of $\beta$ term determines 
  the slope of the (N-Z) dependence so that a larger  $\beta$ 
  gives a steep slope. 
%%  On the other hand, the strength $\alpha$ change the depth of
 %% the proton spin orbit potential . For Z=50 core, only the  proton
%%  g$_{9/2}$  proton orbit dominates  the  spin density $J_p$ in Eq.
 %% (\ref{eq:sd})   so that with negative $\alpha_T$= values
 %%  the spin-orbit splitting  is increased. 
%%  Then energy difference 
%%  $\Delta$ e$(h_{11/2}-g_{7/2})$ is smaller for larger  negative $\alpha_T$. 
%%   Notice that the $\alpha$ gives no isospin dependence on the
%%   spin-orbit potential in general. 

\begin{figure}[htb]
\centering
%\includegraphics{fig_1}% Here is how to import EPS art
%%\rotatebox{-90}{
\includegraphics[width=12cm,clip]{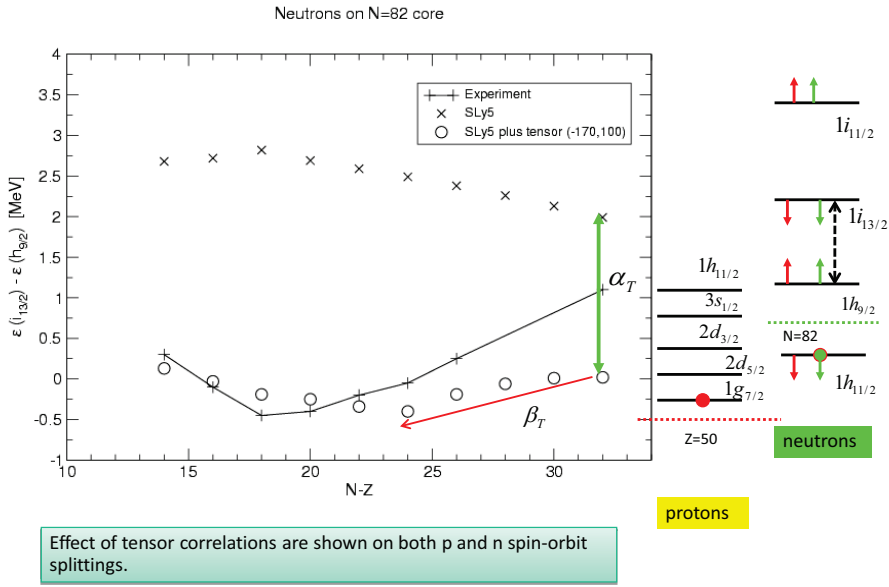}
%%}
%\includegraphics[width=3.2in,height=2.8in]{n82.eps}
\caption{\label{fig:N82}
%\begin{description}
%\item[{\rm Figure 4 :}]
Comparison of energy difference between pairs of
 single-particle 1i$_{13/2}$ and 1h$_{9/2}$ neutron states of N=82 isotopes.
The calculations are performed with and without tensor and central exchange
terms in the spin-orbit potential (\ref{eq:dW}) top of SLy5 parameter set.
The experimental data are taken from ref. \cite{Schiffer}.
See the text for details.
}
\end{figure}

  For the results of Sb-isotones in Fig. \ref{fig:N82}, the  $\Delta$ e$(h_{11/2}-g_{7/2})$ for  N=82 core is plotted as a function of neutron excess.
 Essentially, the same argument can be applied 
  for the neutron energy difference  $\Delta$ e$(i_{13/2}-h_{9/2})$ for 
 N=82 core as
   that for the  $\Delta$ e$(h_{11/2}-g_{7/2})$ for  Z=50 core.
  The last occupied neutrons in $h_{11/2}$ increase the neutron spin-orbit 
splitting so that the $\Delta$ e$(h_{11/2}-g_{7/2})$ become substantially 
smaller by the $\alpha_T$ effect in Eq. (\ref{eq:dW}).
  The isotope dependence is
again explained by the $\beta_T$ effect.  
%  when the proton number increases from the   Z=50 core.
  The 1g$_{7/2}$ and 2d$_{5/2}$ are almost degenerate above the last 
  occupied proton orbit 1g$_{9/2}$ of Z=50 core.
 The two proton 
  orbits,  $d_{5/2}$ and $g_{7/2}$, 
  have opposite  effects on the spin orbit potential (\ref{eq:dW}). 
  The occupation probability is larger for larger $j$ orbit so that 
   1g$_{7/2}$ plays more important role on the spin orbit potential due
  to the tensor interaction  in nuclei with (N-Z)=(32-18) for
   N=82 isotones. Namely the neutron spin orbit splitting is larger for these
  isotones so that the $i_{13/2}$ orbit is down and the $h_{9/2}$ is up.
  These changes make the energy gap $\Delta$ e$(i_{13/2}-h_{9/2})$ smaller
  for the nuclei from (N-Z)= 32($^{132}$Sn) to  (N-Z)=18($^{146}$Gd). 
%%  For nuclei with (N-Z)=(18-14), the proton 2d$_{3/2}$ orbit is gradually 
%%  occupied. This orbit decrease the neutron spin orbit potential so that 
%%   $\Delta$ e$(i_{13/2}-h_{9/2})$ is increasing from  N=18 to N=14.
Thus the role of the triple-even and triplet-odd tensor interactions
are clearly shown in Figs. \ref{fig:Z50} and  \ref{fig:N82} for both 
 proton and neutron spin-orbit splittings.

  The role of the tensor interaction due to the $\beta$ term 
  is essentially expected from the discussion by 
Blatt-Weisskopf \cite{BW} for the deuteron. 
%% In ref.\cite{Otsuka},
%%  the same argument was also presented. 
  The role of
  $\alpha_T$ is new and has not been examined in a quantitative way
 in the mean field calculations 
  since this term comes from the triplet-odd tensor interaction. 
 The  triplet-odd tensor interaction was 
  not included in the studies of refs. \cite{Otsuka,BW}.
  Recently, Brown et al. studied the Skyrme-type tensor interactions
 in $^{132}$Sn and  $^{114}$Sn based on the parameter set SkX.  
  They took both the 
  positive and negative $\alpha_T$ values in the HF calculations and
  concluded that the  negative 
 $\alpha_T$ value gives a better agreement with  the experimental data. 
  This is consistent with the present systematic study of Z=50 isotopes and
  N=82 isotones with HF+BCS model.

\section{Tensor effect on  Gamow-Teller states}

%%rotatebox{-90}{
%%\includegraphics[width=8cm,bb=0.0 0.0 720.0 540.0]{a1s.pdf}
%%}
%%\rotatebox{-90}{
%%\includegraphics[width=8cm,bb=0.0 0.0 720.0 540.0]{a2.pdf}
%%}
%%\end{figure}

 There has been  no RPA or QRPA
(Quasiparticle Random Phase Approximation) program available to
study the effect of the tensor terms on the excited states of
nuclei until very recently.
The tensor terms of the
Skyrme effective interaction was firstly introduced 
 in the self-consistent HF plus RPA
calculations ~\cite{Bai09}, in particular,  in the GT
transitions, which should be affected because of the fact that
the corresponding operator is spin-dependent. In the study of GT
transitions, the quenching problem is of some relevance. The
experimentally observed strength from 10 to 20 MeV excitation energy
(with respect to the ground state of the target nuclei) is about
$50\%$ of the model-independent non-energy weighted sum rule
(NEWSR)
%%; this percentage becomes about $70\%$ if the whole strength
%%in the neighboring energy region is collected~
 \cite{Rapaport}. It would be
very  interesting to study whether the tensor force has an effect in
shifting the strength already at one particle-one hole (1p-1h)
level. Coupling the GT with two particle-two hole states is
essential to describe the resonance width but it is not expected to
affect strongly the position of the main GT peak; the effect of the
tensor force in connection with the 2p-2h coupling was studied in
Ref~\cite{Bertsch}.

It should be noted that $J_q$ gives essentially no contribution
in the spin-saturated cases.
%Consequently, the tensor force
%give no contribution
%to the energy density in both neutron and proton spin-saturated
%nuclei.
Therefore, we choose $^{90}$Zr and $^{208}$Pb as examples to be
calculated. $^{90}$Zr is a proton spin-saturated nucleus, with a
spin-unsaturated neutron orbit $1g_{9/2}$. $^{208}$Pb is chosen as
it is not saturated either in protons or neutrons. The two examples
should allow elucidating separately the role of triplet-even and
triplet-odd terms.

Since the tensor force is spin-dependent and affects the spin-orbit
splitting, the spin mode is very likely to receive strong influence.
%%we study hereafter the GT excitation as the well-known spin mode.
The operator for GT transitions is defined as
\begin{eqnarray}
\hat{O}_{GT\pm}&=&\sum\limits_{im}t^i_{\pm}\sigma_m^i
\end{eqnarray}
in terms of the standard isospin operators,
$t_\pm=\frac{1}{2}(t_x{\pm}it_y)$. In the charge-exchange RPA, the
$t_-$ and $t_+$ channels are coupled and the corresponding
eigenstates emerge from a single diagonalization of the RPA
matrix.

\vspace*{-2cm}
\begin{figure}[htb]
\centering
\includegraphics[width=14cm,clip]{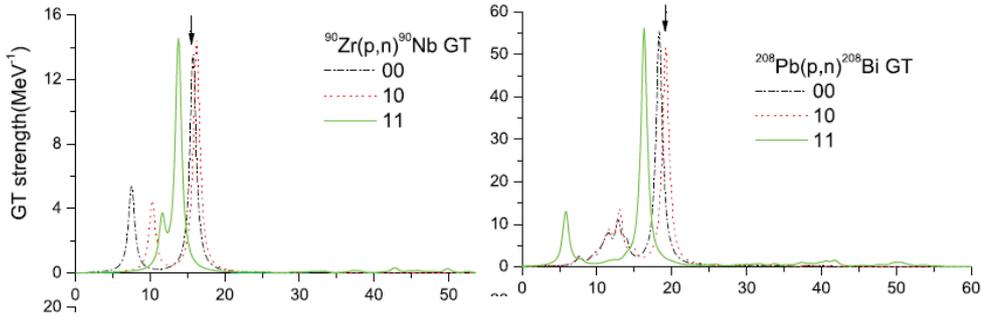}
\vspace*{-3cm}
\caption{\label{fig:208Pb-GT}
%\begin{description}
%\item[{\rm Figure 4 :}]
The GT$_-$ strength in $^{90}$Zr and $^{208}$Pb. The RPA
results are displayed, by smoothing them with Lorentzian function having 1
MeV width. As explained in the text, result labeled by 00
corresponds to neglecting the tensor terms in both HF and RPA; 10
corresponds to including the tensor terms in HF but neglecting them
in RPA; finally, 11 corresponds to including the tensor terms in
both HF and RPA. The arrow denotes the experimental energy.
See the text for details.
}
\end{figure}

The GT$_-$ strength distributions in $^{90}$Zr and $^{208}$Pb are
shown in Fig. \ref{fig:208Pb-GT}. The
calculated results are smoothed by averaging the sharp RPA peaks
with Lorentzian function weighting function having 1 MeV width. The tensor
force affects these results in two ways. Firstly,  it changes the
single-particle energies (s.p.e.) in the HF calculation; secondly,
it contributes to the RPA residual force. We do three different kind
of calculations to analyze separately these effects. In the first
one, the tensor terms are not included at all. In the second one, we
include tensor terms in HF but drop them in RPA. This calculation is
not self-consistent, but it displays the effects of changes in
single-particle energies on the strength distribution. In the last
one, the tensor terms are included both in HF and RPA calculations.
For simplicity, results of the three categories of calculations are
labeled by 00, 10 and 11, respectively.

%\medskip \medskip \medskip \medskip \thinspace \thinspace \thinspace
\begin{table}[t] \centering
\caption{Values of the NEWSR $m_-(0)$ and EWSRs $m_-(1)$ for
$^{90}$Zr and $^{208}$Pb in different excitation energy regions. The
two-body spin-orbit interaction is included in HF but neglected in
RPA calculation. The results labeled by 00 correspond to neglecting
the tensor terms both in HF and RPA; 10 corresponds to including the
tensor terms in HF but neglecting them in RPA; 11 corresponds to
including the tensor terms both in HF and RPA. See the text for a
discussion of the effects of the tensor terms. \label{Table2}}
\begin{tabular}{c|ccccccc}
\hline\hline
& type of     & $m_-(0)$ & $m_-(0)$ & $m_-(1)$ & $m_-(1)$  & $m_-(1)$ & $m_+(1)$ \\
& calculation & 0-30MeV  & 30-60MeV  & 0-30 MeV & 30-60 MeV & total    & total \\
\hline \hline
              & 00 & 29.16 & 0.71 & 395 & 26.2 & 421.8 & 10.1\\
$^{90}$Zr     & 10 & 29.16 & 0.79 & 444 & 22 & 466 & 11.1\\
              & 11 & 27.00 & 2.89 & 366.9 & 122 & 493.2 & 10.3\\
\hline
              & 00 & 127.54 & 3.43 & 2080 & 124.5 & 2212.8 & 18.8\\
$^{208}$Pb    & 10 & 127.38 & 3.68 & 2176 & 93 & 2269 & 21\\
              & 11 & 114.10 & 16.58 &1658 & 694 & 2370 & 19.3\\
\hline
\end{tabular}\thinspace
\end{table}
We have evaluated the amounts of NEWSR $m_-(0)$ and EWSR $m_-(1)$ in
different excitation energy regions, and listed them in
Table~\ref{Table2}.
%The contributions from tensor
%terms to the GT centroid (???) calculated by the EWSR are 2.39 MeV
%and 1.19 MeV
%for $^{90}$Zr and $^{208}$Pb, respectively, which are quite close
%to the analytical values 2.49 MeV and 1.26 MeV. This consistency
%confirms again that the effect of the two-body spin-orbit term
%from RPA correlation is not important compared to that of tensor
%term.
%Unexpectedly,
The EWSR in the energy region below
30 MeV (where the one particle-one hole transitions are located) is
decreased, after the inclusion of the tensor term.
From Table~\ref{Table2}, we also see that an appreciable amount
of EWSR is shifted
from the lower energy region (0-30 MeV) to the higher energy
region (30-60 MeV) by including tensor terms in HF plus RPA
calculations.

We also calculated the values of NEWSR in the 0-30 MeV and 30-60 MeV
energy regions for $^{90}$Zr and $^{208}$Pb. When the tensor is not
included in the residual interaction (i.e., the calculations labeled
by 00 and 10), the values of NEWSR in the energy region between
30-60 MeV for both $^{90}$Zr and $^{208}$Pb are small only few
percent of the NEWSR(Fig, ~\ref{fig:208Pb-GT}. But in the case
11, about $10\%$ of NEWSR is shifted from the lower energy region to
the higher energy region (Corresponding 25\% and 29\%  of EWSR in
$^{90}$Zr and $^{208}$Pb, respectively). Moreover, we can see that
essentially no unperturbed strength appears in this region (see the
Fig. \ref{fig:208Pb-GT}). This means that including tensor terms in
simple RPA calculation shifts about $10\%$ of the GT strength to the
energy region 30-60 MeV. While 2p-2h couplings will increase further
these high energy strength, we would like to stress that the tensor
correlations move substantial GT strength from the low energy region
0-30 MeV to the high energy region 30-60 MeV
 even within the 1p-1h model space.

In $^{90}$Zr,  one can notice that the GT strength is
concentrated in two peaks in the region below 30 MeV. There are only
two important configuration involved which are
$(\pi1g_{9/2}-\nu1g^{-1}_{9/2})$ and
$(\pi1g_{7/2}-\nu1g^{-1}_{9/2})$ (see the left panel of 
 Fig. \ref{fig:208Pb-GT}). When the tensor
term is included only in HF and neglected in RPA, the centroid in
the energy region of 0-30 MeV are moved upwards by about 1.5 MeV,
and the high energy peak at $Ex\sim16MeV$ is moved upwards by only
0.5 MeV, as compared with the results without tensor term. When the
tensor term is included both in HF and RPA, the centroid of the GT
strength in the energy region 0-30 MeV is moved downwards by about 1
MeV, and the high energy peak is moved downwards about 2 MeV, as
compared with the results obtained without tensor term. Including
tensor terms in RPA makes the two main separated peaks closer (this
situation also happens for $^{48}$Ca). This result can be attributed
from the HF and RPA correlations of the tensor term. 
%%From the
% Fromypical effect of the tensor correlations on HF
%field~\cite{Otsuka,Colo}. 
When the $\nu1g_{9/2}$ orbit is filled by
neutrons, the tensor
 correlations give a quenching on the spin$-$orbit splitting between
 $\pi1g_{9/2}$ and  $\pi1g_{7/2}$ orbits
  so that the unperturbed energies of the two
main $p-h$ configurations ~$(\pi1g_{7/2}-\nu1g^{-1}_{9/2})$ and
 $(\pi1g_{9/2}-\nu1g^{-1}_{9/2})$ are closer in energy as is shown
  in Fig. \ref{fig:208Pb-GT}.
The RPA results in Fig. \ref{fig:208Pb-GT} with labeled by (00) and (10) reflect
these changes of HF single particle energies due the tensor
correlations and the energy difference between two peaks is
narrower. 
%%Meanwhile, the RPA correlation of tensor term moves the
%%higher energy peak downward, this effect can be seen in the results
%%in Fig. \ref{fig:90Zr-GT} with labeled by (10) and (11). For GT transitions%% in the
%%energy region of 30-60 MeV, the dominant configuration is
%%$(\pi1g_{7/2}-\nu1g^{-1}_{9/2})$. Its strength is shifted from the
%%low excitation energy region to the high excitation energy region by
%%tensor correlations.

In $^{208}$Pb, from the right panel of 
Fig \ref{fig:208Pb-GT} we see that the GT strength is
concentrated in two peaks in the low energy region of 0-30MeV for
all 00, 10 and 11. There are eleven important configurations which
do contribute to these peaks. When the tensor terms are only
included in HF and neglected in RPA, the centroid of these peaks is
moved upwards about 0.5 MeV, and the higher energy peak at E$_x
\sim$18MeV is also raised by about 0.8 MeV. When  the tensor terms
are included in both HF and RPA calculation, the centroid of these
peak  moves downward by about 1.5 MeV, and the higher energy peak
moves also downwards by about 3.3 MeV, compared with the result
obtained without tensor terms. By including tensor terms in RPA
calculation, the GT strengths in the energy region of 30-60 MeV are
increased substantially by the shift of the strength in the energy
region of 0-30 MeV through the tensor force.

\section{Summary}
We study the effect of tensor correlations on single-particle and 
collective states within Skyrme Hartree-Fock and RPA model.
Firstly, We study the role of tensor interactions in Skyrme effective 
interaction on the isospin dependence of 
spin-orbit splittings in  N=82 isotones and Z=50 isotope.
The different role of the triplet-even and triplet-odd tensor forces 
is elucidated by analyzing the spin-orbit splittings in these nuclei.
The experimental isospin dependence of these splittings cannot be described
  by HF calculations with standard Skyrme forces, 
but is very well accounted for when the tensor forces are introduced.  
%%We conclude that adding the tensor 
%%contribution can  explain in most cases the empirical trends,
%% whereas  this is not the case if one simply employs existing Skyrme
%% parameterizations without the tensor forces~\cite{Colo07}. In conclusion,  We study also the effect of the tensor correlations on
the GT excitations in $^{90}$Zr and $^{208}$Pb in the  HF plus RPA
framework with a Skyrme interaction SIII.
%%If the tensor term is
%%included only in HF but neglected in RPA, the strength distribution
%%is only slightly shifted to higher energy. But 
If the tensor term is included in both HF and RPA,
  the centroid of G-T strength  in
the energy region below 30 MeV is moved downwards by about 1 MeV for
$^{90}$Zr and 3.3 MeV for $^{208}$Pb.  At the same time,
 the dominant  peak at  E$_x \sim$16MeV(18MeV) in $^{90}$Zr($^{208}$Pb)
 is also moved downwards by about 2 MeV(3MeV).
It is pointed out for the first time that about $10\%$ of NEWSR is
moved  in the high energy region of 30-60 MeV by the tensor
correlations in RPA even within $1p-1h$ model space. 
It was pointed out recently that the high energy GT strength 
is shifted by  the coupling between GT ana spin-quadrupole states 
due to  the tensor correlations which has the intrinsic strong coupling
 \cite{Bai09}
%%We have
%%calculated the GT strength by adding presently used tensor terms to
%%$SGII$ and gotten the same result that amount about $10\%$ of the
%%NEWSR appears in the high energy region of 30-60MeV. 
%%N  We give the
%%Nanalytical formula to estimate the effect of the tensor force on the
%%Nmean GT energy. These formulas predict the upwards energy shift of
%%Nthe average excitation energy due to the tensor correlations.  It
%%Nagrees quite well with our numerical RPA results. 
It is interesting
to point out that  the main GT peak, contrarily, gets the energy
shift downward because of the peculiar feature of the tensor
correlations. 
%%Notice that the upward shift of the average energy
%% is the outcome of GT strength appeared in the high energy region 30-60MeV,
%% but not the energy shift of main GT peak.
 The tensor interaction is spin-dependent,
so we expect that it can have important effects not only on the GT
transitions, but on spin-dipole and other spin dependent  excitation
modes as well. 
%%These issues will be discussed in a forthcoming paper.
\vspace*{0.5cm}

%%\begin{acknowledgments}
I would like to thank all the collaborators to proceed the projects of
tensor forces in the mean field models.  Especially 
I would like to thank Gianluca Col\'o for many  stimulating discussions
 in various stages of collaborations.
This work is supported in part by the
Japanese Ministry of Education, Culture ,Sports, Science  and
Technology
  by Grant-in-Aid
for Scientific Research under
 the program number (C (2)) 20540277 .

\end{document}